\documentclass[pra,showpacs,twocolumn,showkeys]{revtex4-1}
\usepackage{graphics}
\usepackage{graphicx}
\usepackage{amsmath}
\newcommand{\bibs}{/Users/jperezmo/Dropbox/References/BibFile}

\begin{document}

\title{Applying universal scaling laws to identify the best molecular design paradigms for third-order nonlinear optics}

\author{Javier Perez-Moreno}
\email{jperezmo@skidmore.edu} \affiliation{Department of Physics, Skidmore College, Saratoga Springs, New York 12866}

\affiliation{Department of Physics and Astronomy, Washington State University, Pullman, Washington 99164-2814}

\author{Shoresh Shafei}
\affiliation{Department of Physics and Astronomy, Washington State University, Pullman, Washington 99164-2814}

\affiliation{Current address: Department of Chemistry, Duke University, Durham, North Carolina 27708-0354}

\author{Mark G. Kuzyk}
\affiliation{Department of Physics and Astronomy, Washington State University, Pullman, Washington 99164-2814}

\begin{abstract}
The scaling of the fundamental limits of the second hyperpolarizability is used to define the intrinsic second hyperpolarizability, which aids in identifying material classes with ultralarge nonlinear-optical response per unit of molecular size.  The intrinsic nonlinear response is a size-independent metric that we apply to comparing classes of molecular homologues, which are made by adding repeat units to extend their lengths.  Several new figures of merit are proposed that quantify not only the intrinsic nonlinear response, but also how the second hyperpolarizability increases with size within a molecular class.   Scaling types can be classified into sub-scaling, nominal scaling that follows the theory of limits, and super-scaling behavior.  Super-scaling homologues that have large intrinsic nonlinearity are the most promising because they efficiently take advantage of increased size.  We apply our approach to data in the literature to identify the best super-scaling molecular paradigms and articulate the important underlying parameters.
\end{abstract}


\maketitle

\section{Introduction}

The quest for materials with enhanced third-order nonlinear-optical response has been fueled by the needs of applications in varied fields such as multi-photon biomedical imaging, photodynamic cancer therapies, optical computing, information transmission, laser technology and real-time holography.  Since the origins of the nonlinear-optical response in organic materials such as dye-doped polymers and van der Waals crystals is found to originate at the molecular level, improving these material's third-order nonlinear properties requires the design and optimization of the substituent nonlinear-optical chromophores.

Marks et al.~have identified new materials paradigms illustrated in small molecules with huge two-photon absorption cross-section,\cite{pati01.01} ultralarge hyperpolarizability,\cite{kang07.01}, and large intensity-dependent refractive index.\cite{He11.01}  Roberts et al. have reported on molecules based on triphenylamine-cored slkynylruthenium dendrimers that have exceptionally large third-order susceptibility.\cite{rober09.01}  Based on the apparent observed improvements from one molecule to another, how can we determine which paradigm has the greatest potential? Is it possible to compare the performance of molecules of vastly different sizes?  Which molecules give the largest nonlinearity per unit of size that scales in a favorable way when the molecule is made larger by adding repeat units?  Can we tell when we have reached a fundamental ceiling?

The theory of quantum limits shows that the strength of the nonlinear optical response of a molecule is bounded. The limit is a function of the number of electrons and the energy difference between the two lowest-energy states, and is reached when the charges are optimally arranged.\cite{kuzyk00.01,kuzyk00.02,kuzyk03.02,kuzyk03.01} A fair comparison of the measure of the performance of a molecule is obtained by comparing its response with others that have the same number of electrons and energy gap.  In practice, molecules of interest have varied energy gaps and electron count, so a more fruitful strategy is to compare the molecular nonlinearity with that of the quantum limit for that number of electrons and energy gap, thus through transitivity, making it possible to compare any two molecules by how well they perform relative to the limits.  This is the approach used here.

The theory of the quantum limits has been used to (1) elucidate the origins of the nonlinear optical response at the molecular level,\cite{tripa04.01, tripa06.01, perez07.02, zhou08.01, perez05.01, perez11.02, perez11.01, perez09.02, van2012dispersion, de12.01} (2) introduce new paradigms for optimization,\cite{perez09.01, perez07.01, perez06.01, perez11.02, Kang05.01, brown08.01, He11.01} and (3) establish fundamental scaling laws.\cite{kuzyk10.01, kuzyk13.01} This paper recognizes that it is not enough to search for the ideal molecule, using previous ones as stepping stones to the ultimate one.  Rather, one must identify a family of molecules that both have a large intrinsic nonlinear-optical response, and which super scales so that the intrinsic nonlinear response grows with size.  This will lead to molecules with ultralarge nonlinear-optical response.  Super scaling is desirable because the absolute second hyperpolarizability grows as a power law greater than an exponent of 3; thus, though fewer large molecules will fit within a fixed volume, their aggregate nonlinear response will be greater than that of many more smaller molecules.  When a molecular class sub scales, the larger response of the larger molecules produces a smaller bulk response.

In this work, we analyze the experimental data in the literature to identify the best molecular candidates for the largest third-order nonlinear optical response. The analysis relies on scaling to determine which molecules are candidates for super scaling, so that longer homologues will become more efficient and approach the quantum limit. This paper is the second part to a companion paper that applies the same principles to the hyperpolarizability.\cite{perez16beta}

The strength of the nonlinear-optical response scales with the size of the quantum system\cite{kuzyk13.01} according to {\em simple scaling} when re-scaling results in a self similar system, as is found for a particle in a box as the walls are moved further apart.  The intrinsic nonlinearity removes this effect so that size is removed from consideration, making comparisons between any two molecules possible.  Data from the literature confirms that most molecules fall into the sub-scaling class, so most present-day design paradigms are based on homologues that are less efficient when they are made larger.  Since most molecules fall far below the fundamental limits, molecules that scale at or less than predictions will become worse as they are made larger.  Even when their absolute nonlinear-optical response is large, their electrons are not being used efficiently and larger homologues will underperform.  Only the molecules that super-scale have potential for reaching the fundamental limit, provided that they have large intrinsic nonlinear-optical response. In this work, we identify existing molecules that super-scale with the goal of identifying structural properties associated with a large nonlinear-optical response that can be applied to making even better materials.

This paper is organized as follows. First we introduce limit theory and scaling, and then propose several figures of merit that apply to a group of homologues.  These figures of merit quantify the type of scaling, the extrapolated molecule size that would yield the fundamental limit, and the magnitude of the nonlinearity at saturation. We apply the figures of merit to a group of molecular classes the most promising systems for applications in third-order nonlinear-optical materials.

\section{Approach}

The molecular property of interest is the second hyperpolarizability, $\gamma$, a fourth rank tensor. Typically the largest component is the diagonal one, so we will focus on the largest component and call it $\gamma$ for simplicity. The fundamental limit of $\gamma$ is calculated using the sum rules and given by:\cite{kuzyk00.02}
\begin{equation}\label{eq:gammaMax}
\gamma_{max}= 4 \left( \frac{e\hbar}{\sqrt{m}} \right)^{4} \frac{N^{2}}{E_{10}^{5}},
\end{equation}
where $e$ and $m$ are the charge and mass of the electron, $\hbar$ is the reduced Plank constant, $N$ is the effective number of electrons, and $E_{10}$ is the energy difference between the first and the ground state. Using $esu$ units we can approximate Equation \ref{eq:gammaMax} as:
\begin{equation}\label{eq:gammaMaxUnits}
\gamma_{max} \left[\frac{cm^6}{erg}\right] = 29,700 \times 10^{-36} N^2/E_{10}\left[eV \right]^5 ,
\end{equation}
where the quantities in brackets are the units. The conversion between energy of a photon in $eV$ and its associated wavelength $\lambda$ in nanometers is $\lambda[nm] = 1240/E[eV]$.

The fundamental limits define an absolute maximum, so the ratio of the measured nonlinearity to the limit is a dimensionless parameter of magnitude less than unity.  The intrinsic second hyperpolarizability is defined as the ratio\cite{watki12.01}
\begin{equation}\label{eq:gammaInt}
\gamma_{int} = \frac {\gamma} {\gamma_{max}}.
\end{equation}
It has be shown that in general, the second hyperpolarizability scales in the same manner as the fundamental limit, or\citep{kuzyk13.01}
\begin{equation}
\label{eq:gammascales}
\gamma \propto \frac{N^{2}}{E_{10}^{5}}.
\end{equation}
This kind of scaling, which is obeyed by all self-similar structures, is called {\em ``simple scaling''}.  The ratio defined by Equation \ref{eq:gammaInt} eliminates simple scaling, and is thus said to be scale invariant or size independent.

The Schr\"{o}dinger Equation is invariant under transformations in which the lengths are re-scaled by a factor $\epsilon$ if the energies are simultaneously re-scaled by a factor $1/\epsilon^2$.\cite{zhou08.01,kuzyk10.01} Such re-scaling would change the absolute value of the second hyperpolarizablity but would leave the intrinsic second hyperpolarizability unchanged. This idea applies to molecules, so we assess the scaling behavior of molecular classes using the change in intrinsic second hyperpolarizability as a function of size as a metric.  A molecular {\em ``class''} is a collection of homologue molecules of varying sizes.  When the intrinsic second hyperpolarizability in a class is independent of the size, the class is assigned to the scaling type calked simple scaling. If the intrinsic second hyperpolarizability increases or decreases with size the class is assigned to the super-scaling or sub scaling type, respectively.  The target paradigm is a molecular class with a large nonlinear response that super scales. In this paper we use these concepts to define figures of merit and identify the best molecular paradigms.

\section{Results and discussions}

\begin{figure*}
\centering
  \includegraphics[width=4in]{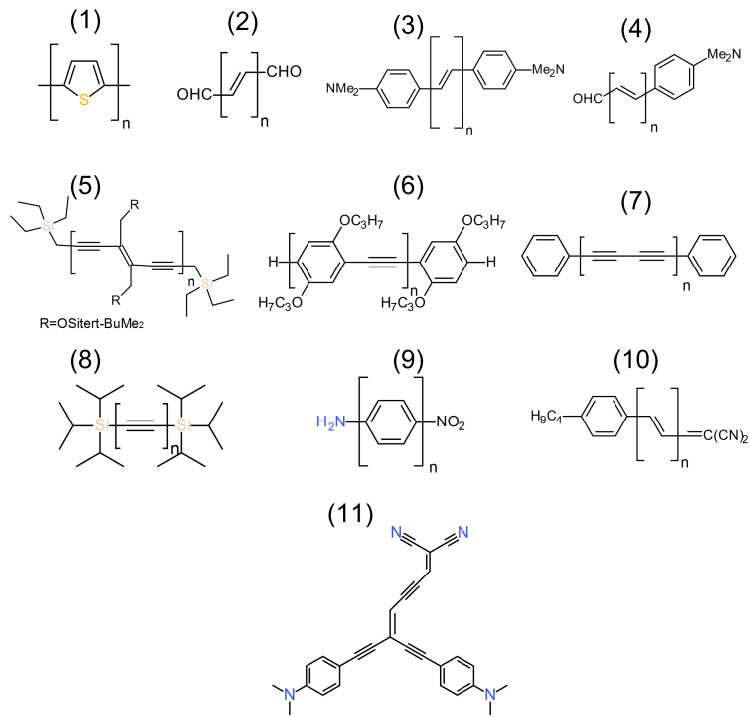}\\
  \caption{The molecular classes considered in this work.}\label{fig:gamma}
\end{figure*}

Figure \ref{fig:gamma} shows the molecular classes whose second hyperpolarizabilities are studied. In each case, the base molecule is shown, and the class is defined by varying the number of repeat units, $n$. The calculation of the intrinsic second hyperpolarizability requires as an input the measured second hyperpolarizability, the effective number of electrons, $N$, and the energy difference between the ground and first electronic excited states, $E_{10}$ so that Equation \ref{eq:gammaInt} can be evaluated using Equation \ref{eq:gammaMax}.

The absolute second hyperpolarizabilities are determined from measurements reported in the literature, the energy difference $E_{10}$ is determined from the wavelength of maximum absorption, and the effective number of electrons is determined according to:
\begin{equation}\label{eq:gammaEffN}
N_{\gamma} =  \left( \sum_{i} N_i^2 \right)^{1/2},
\end{equation}
where the sum is over each contiguous conjugated path.\cite{kuzyk03.03}  For a single conjugated path, there are two electrons per double or triple bounds, and the effective number of electrons is simply the total number of $\pi$-electrons in the conjugated path. The number of effective electrons is calculated using Equations \ref{eq:gammaMax}. The values of $E_{10}$ and $N$ for all the molecular classes are tabulated in Table \ref{tab:gamma}.

\begin{figure}
  \centering
  \includegraphics{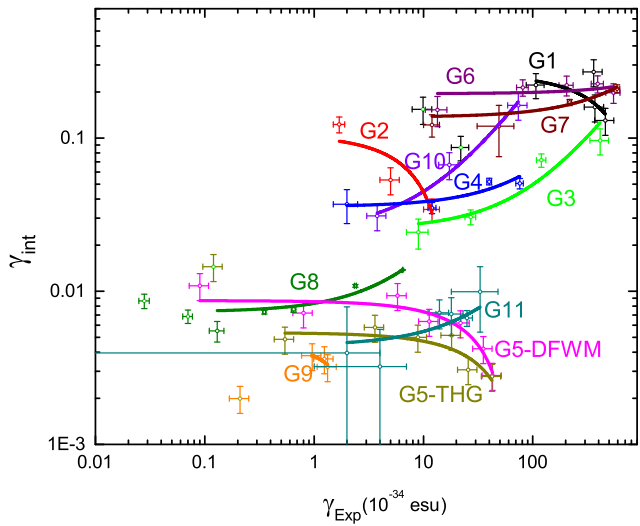}\\
  \caption{Plot of the intrinsic second hyperpolarizability $\gamma_{int}$ as a function of the measured  absolute second hyperpolarizability $\gamma_{exp}$. The fit is to the function $\gamma_{int}=c \gamma_{exp} + d$ and the fitting parameters are given in Table \ref{tab:gamma}. Since the scales for the horizontal and vertical axis are logarithmic, the linear fits appear as curves.}\label{fig:GammaIntExp}
\end{figure}

Figure \ref{fig:GammaIntExp} plots the intrinsic second hyperpolarizability as a function of the absolute second hyperpolarizability. While the absolute second hyperpolarizability spans 4 orders of magnitude, the intrinsic hyperpolarizability spans only two orders of magnitudes. As in the case of the first hyperpolarizability, this is an indication that most of the measured variations are due to simple scaling. An understanding of the origin of the two orders of magnitude variation of the intrinsic second hyperpolarizability could be used to make molecules with better scaling that would translate into much bigger absolute nonlinearities. It is interesting to note that the longest molecule in class G5 has a respectable second hyperpolarizability that is larger than most of the other molecules; but, its intrinsic nonlinear-optical response is the smallest of all the molecules.

\begin{figure*}
  \centering
  \includegraphics[width=4.5in]{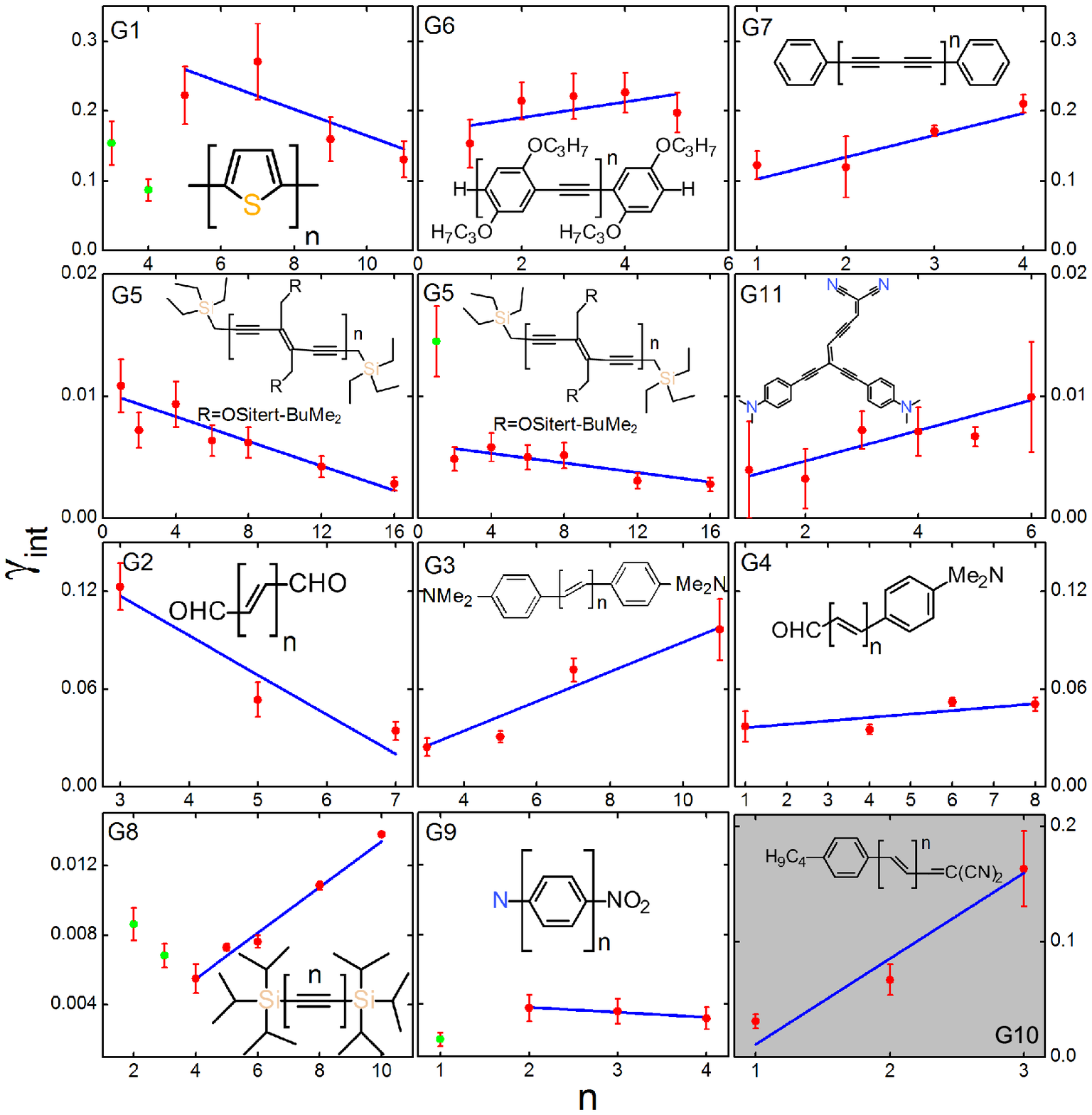}\\
  \caption{A plot of $\gamma_{int}$ as a function of the number of repeat units, $n$. The fit function is $\gamma_{int} = a n + b$ and the fit parameters are given in Table \ref{tab:gamma}. The green points are excluded from the linear fits. Notice that the vertical scale is the same for each row with the exception of series G10, which is shown shaded.}\label{Gmerger}
\end{figure*}

Figure \ref{Gmerger} plots the intrinsic second hyperpolarizability (red points) as a function of the number of repeat units for each molecule class. As it was found for the first hyperpolarizability, the relationship is approximately linear. The blue lines show the liner fit ($\gamma_{int}= a \cdot n + b$). The slope of the line determines the nature of scaling in each series. In some cases, such as series G8, G9 and G1, the effect of the molecular ends is large for the shortest molecules. In these cases, the shorter members in the series, shown as green points, are excluded from the linear fits. The fit parameters $a$ and $b$ are listed in Table \ref{tab:gamma}, together with the values of $c$ and $d$ which are determined from the fit $\gamma_{int} = c \gamma_{exp} + d$, as shown in Figure \ref{fig:GammaIntExp}. Other parameters in Table \ref{tab:gamma} are discussed later.  Notice that $a$, which quantifies the degree of scaling, is also the incremental intrinsic second hyperpolarizability per repeat unit, or
\begin{equation}\label{eq:a-gamma}
a = \frac {\partial \gamma_{int}} {\partial n},
\end{equation}
and $b$ is the extrapolated value of the second hyperpolarizability in the limit of zero repeat units:
\begin{equation}\label{eq:b-gamma}
b = \left. \gamma_{int} \right|_{n=0}.
\end{equation}

\begin{table*}
\footnotesize
\caption{The scaling parameters for the second hyperpolarizability molecular classes. The listed value of $E_{10}$ is for the molecule in the class with the least number of repeat units, denoted by $n^{\prime}$. $N$ is the number of effective electrons of the molecule with the least number of repeat units. The fitting functions are $\gamma_{int} = a n + b$ and $\gamma_{int} = c \gamma_{exp} + d$. $\gamma_5^{T}$ and $\gamma_5^{D}$ are second hyperpolarizabilities measured by third harmonic generation and degenerate four wave mixing. $\gamma_{int}^{max}$ is the value of the best intrinsic hyperpolarizability in the class, for the molecule that has $n_{max}$ repeat units. $\gamma_{n=1}$ is the value of the absolute first hyperpolarizability for the molecule with $n=1$ repeat units in the series. $\gamma_{SAT}$ is the predicted value of the absolute second hyperpolarizablity at the saturation length, $FOM_{\gamma}$ is the proposed figure of merit (Equation \ref{eq:gammafom}) and $\Delta \gamma_{exp}$ is the incremental addition to the absolute second hyperpolarizability per repeat unit (Equation \ref{eq:gammarac}).}\label{tab:gamma}
  \centering
  \begin{tabular}{|c | c | c | c | c | c | c | c | c | c | c | c | c | c |}\hline
 Class & $E_{10}$ & ~$n^{\prime}$~ & $N$ & $a$ & $b$ & $\gamma_{max}^{int}$ & ~$n_{max}$ & $c$ & $d$ & $\gamma_(n=1)$ & $\gamma_{SAT}$ & $FOM_{\gamma}$  & $\Delta \gamma_{exp}$ \\
    & (eV) & & & $\times 10^{-3}$ & $\times 10^{-3}$ & $\times 10^{-3}$ & & $\times 10^{30} esu^{-1}$ & $\times 10^{-3}$ & $\times 10^{-34} $ & $\times 10^{-34} $ & $\times 10^{-34} $ & $\times 10^{-34} $ \\
    & & & & & & & & & & esu & esu & esu & esu \\ \hline
  $\gamma_1$ \cite{thien90.01}& 3.67 & 3 & 4 & -19 $\pm$ 13 & 355 $\pm$ 99 & 271 $\pm$ 55 &7 & -1.0 $\pm$ 0.3 & 253 $\pm$ 95 & 9.9 (n=3) & - & - & - \\
  $\gamma_2$ \cite{pucce93.01}& 3.780 & 3 & 2 & -24 $\pm$ 7 & 190 $\pm$ 36 & 123 $\pm$ 14 &3& -90 $\pm$ 50 & 124 $\pm$ 32 & 1.7 (n=3)& - & - & - \\
  $\gamma_3$ \cite{pucce93.01}& 3.039 & 3 & 14 & 9 $\pm$ 1 & -2 $\pm$ 13 & 96 $\pm$ 19 &11& 2.0 $\pm$ 0.4 & 350 $\pm$ 12 & 9 (n=3) & 3250 & 29 & $4.5 \pm 1.4$ \\
  $\gamma_4$ \cite{pucce93.01}& 3.229 & 1 & 8 & 2.0 $\pm$ 0.9 & 34 $\pm$ 4 & 52 $\pm$ 3 &6& 2.0 $\pm$ 0.8 & 36 $\pm$ 3 & 2 & 4820 & 10 & $1.0 \pm 0.9 $ \\
  $\gamma_5^{T}$ \cite{guble99.01}& 4.189 & 1 & 6 & -0.20 $\pm$ 0.07 & 6.0 $\pm$ 0.5 & 15 $\pm$ 3 &1& -0.7 $\pm$ 0.2 & 6.0 $\pm $ 0.4 & 0.12& - & - & - \\
  $\gamma_5^{D}$ \cite{guble99.01}& 4.189 & 1 & 6 & -0.5 $\pm$ 0.1 & 1.3 $\pm$ 0.8 & 10 $\pm$ 2 &1 & -2.0 $\pm$ 0.4 & 10.0 $\pm$ 0.7 & 0.09 & - & - & - \\
  $\gamma_6$\cite{meier01.01} & 3.669 & 1 & 14 & 11 $\pm$ 9 & 168 $\pm$ 30 & 226 $\pm$ 29 &1 & 0.6 $\pm$ 0.7 & 185 $\pm$ 23 & 13.4 & 13583 & 180 & $20 \pm 40 $ \\
  $\gamma_7$\cite{luu05.01} & 3.780 & 1 & 16 & 32 $\pm$ 11 & 71 $\pm$ 27& 210 $\pm$ 13 &7 & 2.0 $\pm$ 0.3 & 116 $\pm$ 6 & 12 & 4420 & 152 & $16 \pm 8 $ \\
  $\gamma_8$\cite{eisle05.01} & 6.2 & 2 & 2 & 1.0 $\pm$ 0.1 & 0.2 $\pm$ 0.6 & 14..0 $\pm$ 0.06 &10 & 20 $\pm$ 4 & 6.0 $\pm$ 0.6 & 0.028 (n=2) & 4970 & 5 & $0.05 \pm 0.02 $ \\
  $\gamma_9$\cite{meier05.01} & 3.324 & 1 & 8 & -0.30 $\pm$ 0.07 & 4.0 $\pm$ 0.2 & 4.0 $\pm$ 0.8 &2 & -10 $\pm$ 7 & 5.0 $\pm$ 0.9 & 0.21 & - & - & -\\
  $\gamma_{10}$\cite{meier05.01} & 3.324 & 1 & 14 & 74 $\pm$ 17 & -63 $\pm$ 45 & 163 $\pm$ 33 &3 & 4.0 $\pm$ 0.6 &-35 $\pm$ 22 & 3.78 & 2587 & 185 & $20 \pm 7$ \\
  $\gamma_{11}$\cite{May07.01} & 2.857 & 1 & 26 & 1.0 $\pm$ 0.2 & 2.0 $\pm$ 0.9 & 10 $\pm$ 6 & 6 & 20 $\pm$ 300 & 3.00 $\pm$ 0.05 & 4 (n=2) & 498 & 0.5 & $0.05 \pm 0.8$ \\
  \hline
\end{tabular}
\end{table*}

\begin{figure}
  \centering
  \includegraphics{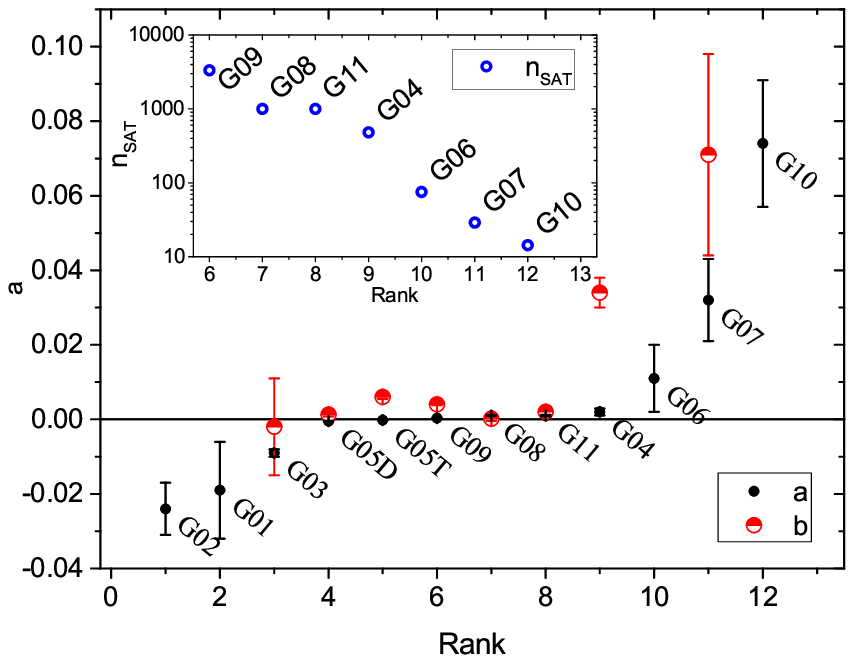}\\
  \caption{Plot of the incremental intrinsic second hyperpolarizability per repeat unit $a$  and the intrinsic second hyperpolarizability of the base $b$, i.e. when $n=0$. The classes are ranked based on the value of $\gamma_{int}^{max}$ (listed in Table \ref{tab:gamma}). The inset shows $n_{SAT}$, the number of repeat units required to attain $\gamma_{int}=1$ assuming that scaling remains linear.}\label{fig:a-and-b-gamma}
\end{figure}

A plot of $a$ and $b$ for all the molecular classes is shown in Figure \ref{fig:a-and-b-gamma}. On the horizontal axis, the classes are ranked based on the value of $\gamma_{int}^{max}$ (listed in Table \ref{tab:gamma}).  The inset shows the number of repeat units required to attain $\beta_{int}=1$ at the saturation length $n_{SAT}$ assuming that scaling remains linear.

The scaling type of a molecular class is given by the sign of $a$.  Classes G06, G07, and G10 super-scale ($a>0$) while classes G04, G08, and G11 show nominal scaling ($a \approx 0$) within experimentally uncertainty. Classes G01, G02, and G03 sub-scale ($a<0$) while classes G05 and G09 sub-scale but are within experimental uncertainty of nominal scaling. Classes G06, G07 and G10 also require the minimum number of repeat units of all to reach saturation, when $\gamma_{int} = 1$. Class G10 is the best of all, with $n_{sat}$ between 10 and 20, a synthetically-achievable target. However, G10 has the fewest number of data points, so extrapolation may be inaccurate.

\begin{figure}
  \centering
  \includegraphics[width=3.4in]{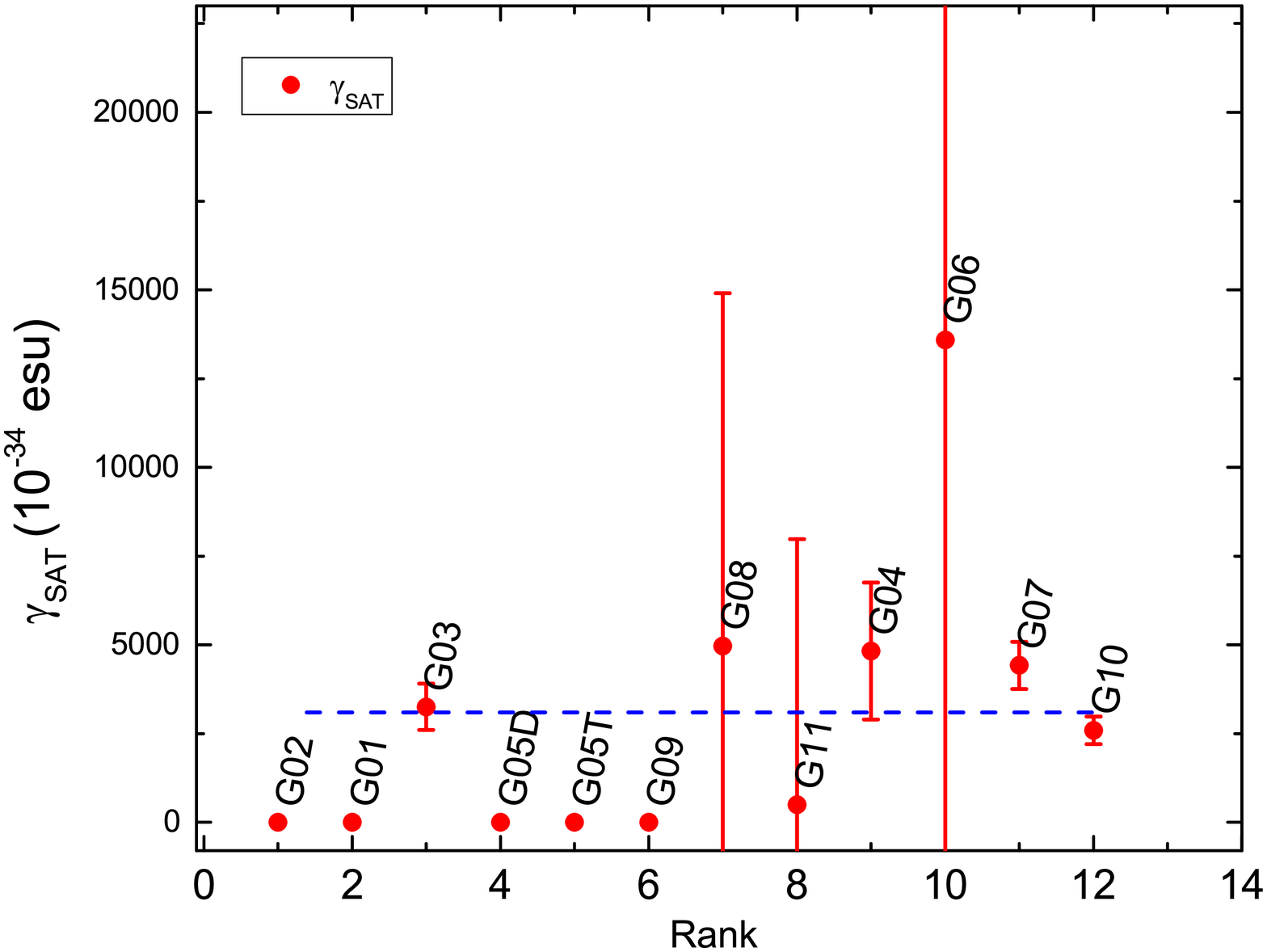}\\
  \caption{Plot of $\gamma_{SAT}$, the absolute value of the second hyperpolarizability at saturation, defined by $\gamma_{int} = 1$, as a function of rank. The classes are ranked based on their highest value of $\gamma_{int}$, which is labelled $\gamma_{int}^{max}$ in Table \ref{tab:gamma}.}\label{fig:gammaSAT}
\end{figure}

A possible figure of merit is the number of repeat units required to attain the quantum limit (such as $\gamma_{int} \rightarrow 1$) and the nonlinear response saturates. The number of repeat units required to saturate the absolute second hyperpolarizability, $n_{SAT}$ is obtained by extrapolation of the linear fit $\gamma_{int} = a n + b$:
\begin{equation}
n_{SAT}=\frac{1-b}{a}.
\end{equation}
The smaller the value of $n_{SAT}$, the better the molecular class. Figure \ref{fig:gammaSAT} shows $\gamma_{SAT}$, the predicted absolute value of the second hyperpolarizability when the number of repeat units is large enough to attain the quantum limit as a function of rank. The ranking is based on the highest value of $\gamma_{int}$ in the class, $\gamma_{int}^{max}$ (listed int Table \ref{tab:gamma}). Interestingly, all classes that super-scale have the same saturating second hyperpolarizability within experimental uncertainty.

\begin{figure}
  \centering
  \includegraphics[width=3.4in]{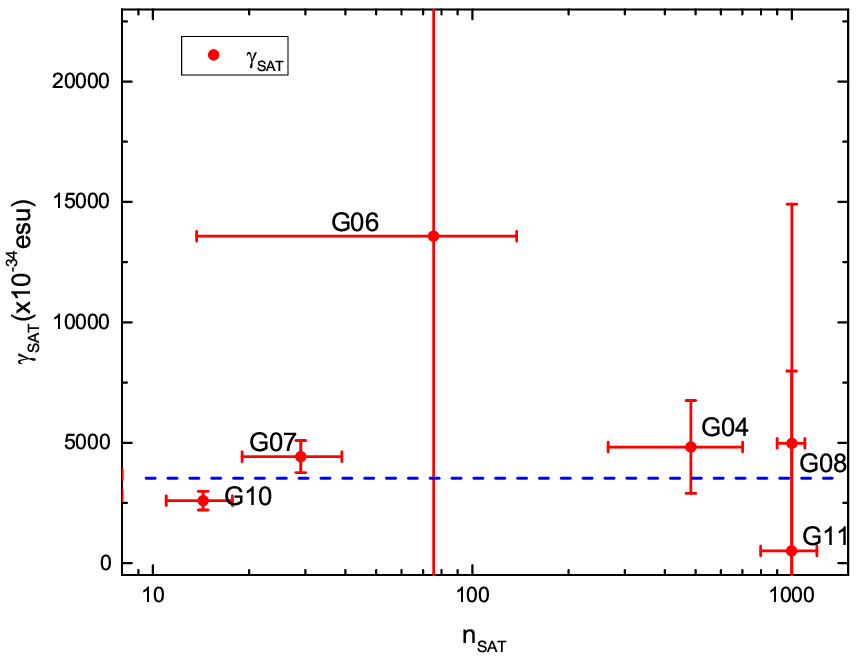}\\
  \caption{The expected absolute second hyperpolarizability that would be achieved when the class saturates ($\gamma_{SAT}$), as a function of the number of repeat units needed to reach saturation ($n_{SAT}$). The expected absolute hyperpolarizability for classes that sub-scale (not shown in the plot) is zero.}\label{fig:gammaVnSAT}
\end{figure}

Figure \ref{fig:gammaVnSAT} shows a plot of $\gamma_{SAT}$ as a function of $n_{SAT}$. Since all of the super-scaling series have about the same values of $\gamma_{SAT}$, the ones with the smallest value of $n_{SAT}$ are best.

As homologues are made longer, it becomes more unlikely that the molecules will retain the scaling properties due to breaks in conjugation. A more practical measure of the scaling performance of a class takes into account the number of repeat units needed to attain the largest value allowed for the second hyperpolarizability. A figure of merit that accounts for both $\gamma_{SAT}$ and $n_{SAT}$ is,
\begin{equation}
FOM_{\gamma} =  \frac {\gamma_{SAT}} {n_{SAT}}.
\label{eq:gammafom}
\end{equation}

\begin{figure}
  \centering
  \includegraphics[width=3.4in]{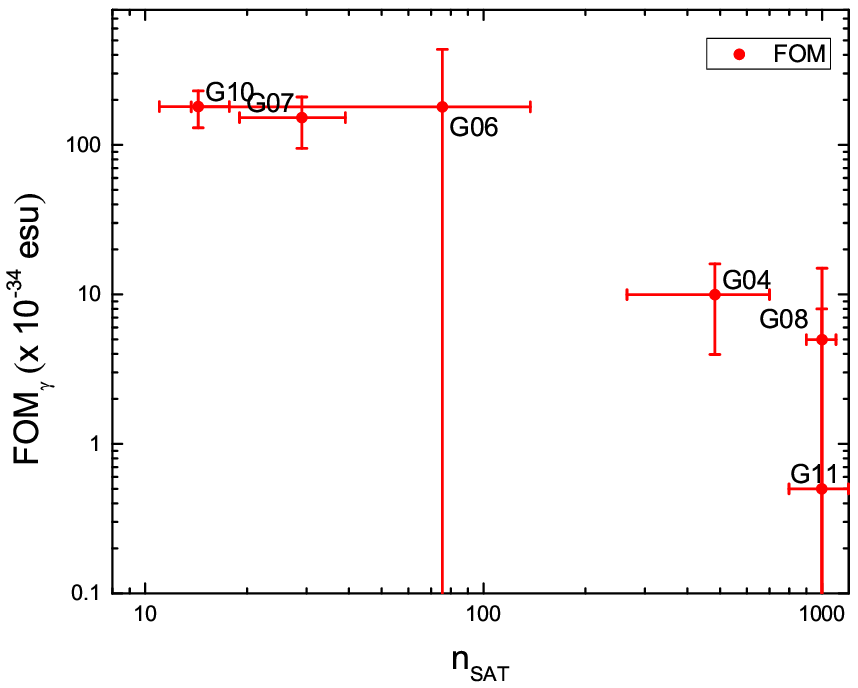}\\
  \caption{The figure of merit ($FOM_{\gamma}$) defined as the ratio $\gamma_{SAT} / n_{SAT}$ as a function of the number of repeat units needed to reach saturation.}\label{fig:gammaFOM}
\end{figure}

Figure \ref{fig:gammaFOM} shows the  figure of merit ($FOM_{\gamma}$) as a function of the number of repeat units needed to reach saturation, $n_{SAT}$. The super-scaling classes all share a very similar value of $FOM_{\gamma} \approx 170 \times 10^{-34} $ esu.

\begin{figure}
  \centering
  \includegraphics[width=3.4in]{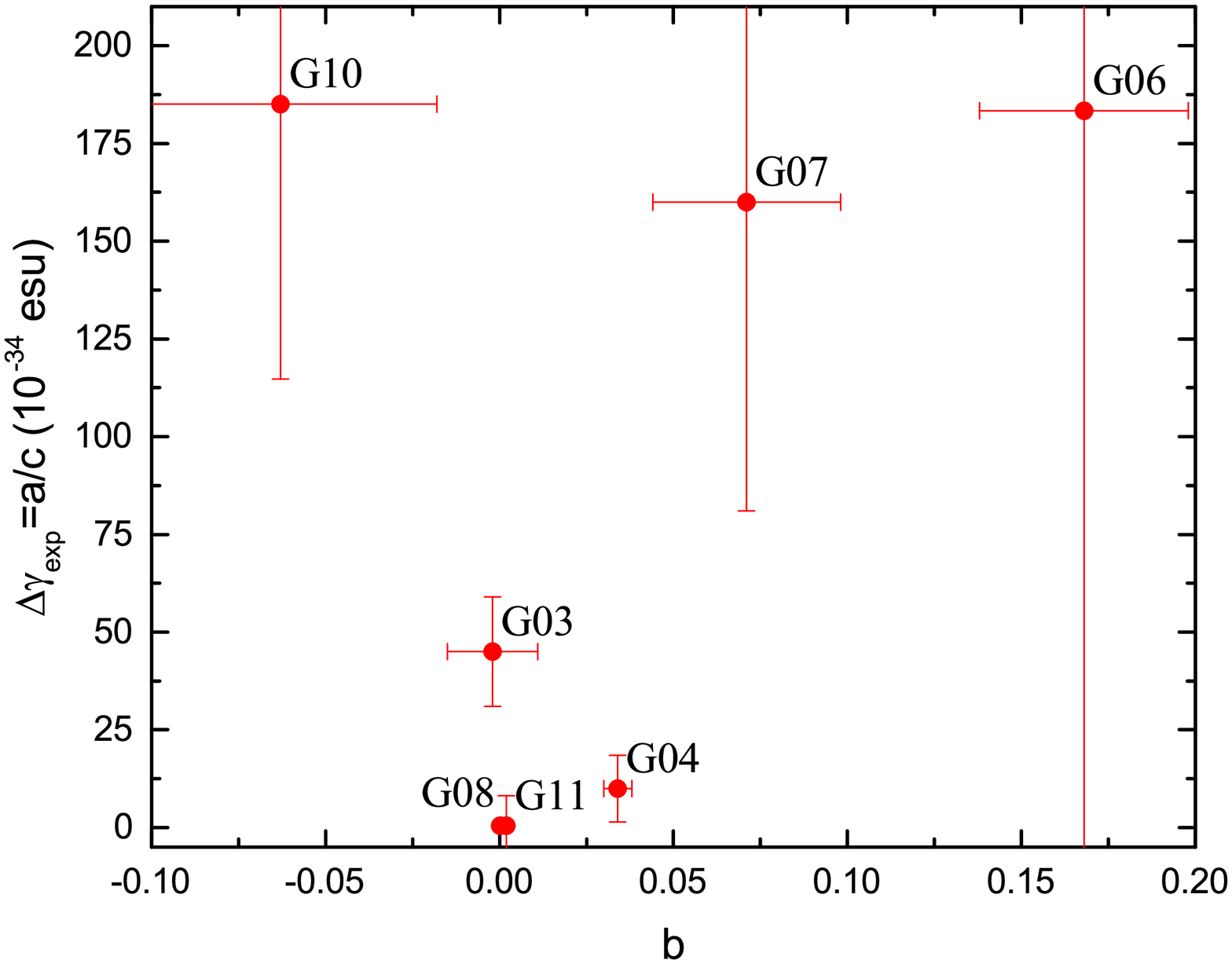}\\
  \caption{The incremental contribution to the absolute second hyperpolarizability per repeat unit, $\Delta \gamma_{exp}$, as a function of $b$ (the intrinsc hyperpolarizability in the limit of the base molecule, i.e. with $n=0$).}\label{fig:GammaPer}
\end{figure}

A more telling quantity, when the goal is to make just a few longer molecules, is how much the absolute second hyperpolarizability increases as a new repeat unit is added and is parameterized by $\Delta \gamma_{exp}$, which can be expressed as:
\begin{equation}
\Delta \gamma_{exp} = \frac{a}{c}.
\label{eq:gammarac}
\end{equation}
The values of $\Delta \gamma_{exp}$ are plotted in Figure \ref{fig:GammaPer} and listed in Table \ref{tab:gamma}. Within experimental uncertainty, each of the super-scaling molecules have the same incremental contribution ($\Delta \gamma_{exp} \approx 180 \times 10^{-34}$ esu). In turn, the nominal scaling classes all share a similar value ($\Delta \gamma_{exp} \approx 5 \times 10^{-34}$ esu).

Classes G07 and G10 have both the largest figure of merit and the highest incremental contribution to the absolute second hyperpolarizability per repeat unit. While class G06 appears to be in line with the others, its experimental uncertainty is high, so its its figure of merit could actually be low. The data in Figure \ref{Gmerger} shows that the error bars are larger than the slope, and that nominal scaling is also consistent with the data. Similarly, its value of $\Delta \gamma_{exp}$ could actually be null or even negative. Thus, enough data is not available to evaluate this class, while classes G07 and G10 deserve further discussion. However, given that only three points were used to determine the nature of scaling in class G10, additional measurements are needed to confirm that this system is in the super scaling class.

The data for class G07 is the most reliable. As was found for the first hyperpolarizability, the simplest bridge forming a linear chain appears to best. In contrast, classes G01 and G09 - with a more complex bridge, do not have good scaling properties. Also note that the cyclic end groups found on class G07 seem to make the polyyne bridge more effective, which is clear when comparing it to class G08, which shares the same bridge but has non-conjugated end groups. As a result, class G08 scales well but its absolute second hyperpolarizability is small, requiring many more repeat units to reach $\gamma_{SAT}$.

Note that classes G03 and G04 have the cyclic end group theme, thus leading to a good saturation second hyperpolarizability; but, the polyene bridge does not seem as effective as the polyyne one. We thus conclude that polyyne bridges with cyclic conjugated end groups may be the best paradigm where the end groups are sources and sinks of charge and the polyyne bridge serves as an efficient conduit between the two. Indeed, materials based on polyynes have been studied by Slepkov and coworkers\cite{slepk04.01, luu05.01, eisle05.01} as examples of systems that scale well. Molecular classes such as those studies by May et al., of the type given by G11\cite{May05.01, May07.01} were found to have very large second hyperpolarizabilities for relatively small molecules, though G11 may suffer somewhat in its scaling properties because the chains are not aligned to reinforce the nonlinear response.

\section{Approach to analyzing molecular series}

The importance of the present work is no sot much in the results that we have presented, which serve as an example, but in the protocols that we define for a methodology that identifies the promising series of molecules for further study and optimization for scale-up. Based on the examples above, we propose the following approach.

The goal is to find the ideal unit that can be scaled up by linking the units together. The simplest units are ones that connect to form linear chains, but others are possible, including the formation of dendrimers, space filling structures, or any other novel shapes. The units can be stand alone; used to link two ends together -- thus having the end type as an additional degree of freedom; or can be formed into fractal-like dendrimer units with multiple external units and joints.

The evaluation protocol proceeds as follows:

\begin{enumerate}

\item \label{step-Type}
Identify a structure type that includes repeat units and end/exterior units that is expected to show promise based on semi-empirical calculations or intuition

\item \label{step-ChooseEnds}
Choose end/exterior units and keep them fixed and synthesize a series of structure of varying length between 1 and a minimum of 4 repeat units.

\item
Measure the linear absorption spectrum for each to determine $\gamma_{max}$, and then measure $\gamma$ as a function of the number of repeat units to determine $\gamma_{int}$.

\item
From a linear fit of $\gamma_{int}$ versus $\gamma_{Exp}$ and $\gamma_{int}$ versus $n$, determine $n_{SAT}$, $\gamma_{SAT}$ and the Figures of merit.

\item
If the figure of merit for $\gamma$ exceeds $10^{-32} esu$, then make structures with greater numbers of repeat units. Otherwise, go to Set \#\ref{step-ChooseEnds}.

\item
If the scaling law breaks down for longer units, start again at Step \#\ref{step-Type}.  If not, you have a promising molecule for ultra-large second hyperpolarizability.

\end{enumerate}

This procedure identifies useful paradigms that have the potential for ultra-large third-order nonlinear-optical response. Since making lager molecules is a more involved process, the proposed methodology identifies a series that is worth the effort for additional synthetic efforts.

\section{Conclusion}

Making a direct comparison between the nonlinear-optical response of two molecules is problematic because they may be of differing sizes, so differences may be due solely to simple scaling and not to the intrinsic nonlinear response of the molecule. The size of a molecule is not well defined from the quantum perspective because molecules do not have sharp boundaries. However, the difference in energy between the first excited state and the ground state, $E_{10}$, and the effective number of electrons, $N$, defines a size, which is embodied in the fundamental limit of the second-order nonlinear response, $\gamma_{max}$, a function of only $N$ and $E_{10}$. Dividing the nonlinear response by the fundamental limit defines the intrinsic response, which is a scale invariant property that can be used to compare molecules of disparately different sizes. Indeed, the range of the intrinsic nonlinear is much smaller than the absolute nonlinearities because much of the difference is due to size effects.

Using the idea of scale invariance, we have defined a figure of merit that can be used to compare a series of molecules that differ mostly in just their length. This figure of merit can be used to identify new paradigms that are scalable; that is, longer versions of the molecule return a nonlinearity that is far larger than one would attain if it were due only to the increased length.

We have shown how this method can be used to analyze which material classes are the most promising. In the case of the second hyperpolarizability, we find that that the response is optimized by the simple polyyne bridge with simple cyclic end groups.

More importantly, our work uses a review of the literature to illustrate a new approach for identifying better molecular classes. Using this type of a well-defined procedure may be required to make the next big leap in the design of new molecules.

\section{Funding Information}

We acknowledge the National Science Foundation(ECCS-1128076) for generously supporting this work.

% Bibliography
\bibliography{\bibs}

\end{document}